\documentclass[11pt, a4paper]{article}
\usepackage{jcappub}
\usepackage{amsmath}
\usepackage{graphicx,epsf,hyperref}
\usepackage{color,soul}
\usepackage{xcolor}

\def\be{\begin{equation}}
\def\ee{\end{equation}}
\def\bea{\begin{eqnarray}}
\def\eea{\end{eqnarray}}
\def\lb{\left(}
\def\rb{\right)}
\def\lbs{\left[}
\def\rbs{\right]}
\def\lbc{\left\{}
\def\rbc{\right\}}

\def\n{\nabla}

\def\L{{\pounds}}

\def\dr{{\delta\rho}}
\def\H{{\cal H}}
\def\cs2{c_{\rm{s}}^2}
\def\U0{{\bar U_0}}
\def\wt{\widetilde}
\def\12{\frac{1}{2}}

\def\J{{J}}
\def\A{{A}}
\def\V{{V}}
\def\W{{W}}
\def\U{{\Upsilon}}

\def\X{{\cal{X}}}
\def\Xv{{{\cal{X}}_{\rm{v}}}}

\title{Vector and tensor contributions to the curvature perturbation at second order}
\author{Pedro Carrilho}
\author{and Karim A. Malik}

\affiliation{Astronomy Unit, School of Physics and Astronomy, Queen Mary University of London, Mile End Road, London, E1 4NS, UK}

\emailAdd{p.gregoriocarrilho@qmul.ac.uk}
\emailAdd{k.malik@qmul.ac.uk}

\date\today

\abstract{
We derive the evolution equation for the second order curvature
perturbation using standard techniques of cosmological perturbation
theory. We do this for different definitions of the gauge invariant curvature perturbation, arising from different splits of the spatial metric, and compare the expressions. The results are valid at all scales and include all
contributions from scalar, vector and tensor perturbations, as well as
anisotropic stress, with all our results written purely in terms of gauge
invariant quantities. Taking the large-scale approximation, we find
that a conserved quantity exists only if, in addition to the
non-adiabatic pressure, the transverse traceless part of the
anisotropic stress tensor is also negligible. We also find that the version of the gauge invariant curvature perturbation which is exactly conserved is the one defined with the determinant of the spatial part of the inverse metric.}


\dedicated{Published in JCAP as \href{http://dx.doi.org/10.1088/1475-7516/2016/02/021}{JCAP 02 (2016) 021}}

\arxivnumber{1507.06922}

\begin{document}
\maketitle

\section{Introduction}
Advances in observational cosmology in the last decades have provided
us with countless ways to test cosmological models and so far,
the \emph{Standard Model of Cosmology} has passed all of them. In the
simplest version this standard model of cosmology comprises a period
of accelerated expansion, inflation, at the very beginning and has
the dynamics of the universe dominated by dark matter, and later dark
energy, after an initial epoch of radiation domination.

In recent years the results from the WMAP \cite{WMAP} and the
Planck \cite{Planck} missions have provided us with an unprecedented
amount of information from the temperature anisotropies in the Cosmic
Microwave Background (CMB), constraining the allowed parameter
space. Also, recent B-mode polarisation measurements \cite{Bmode} have
put additional tight bounds on the tensor-to-scalar ratio.
In spite of all this progress, the correct model of the early Universe
is still unknown, as there are many inflationary potentials that are
consistent with the experimental results. Future surveys will provide
an even further improvement in the experimental accuracy and should
probe more scales than ever before. However, in order to make the most
of these new observations, further efforts are required also on the
theory side, for example to compute observables with the precision
required by the experiments.

According to inflationary theory, the seeds of structure were
generated during inflation through the quantum fluctuations of one or
more light scalar fields. To study the evolution of these initial
perturbations we use Cosmological Perturbation Theory. This amounts to
adding inhomogeneous perturbations to the
Friedmann-Lema\^itre-Robertson-Walker (FLRW) solution and studying
their evolution equations. While the linear theory seems to have been
successful so far, higher order corrections seem necessary, in order
to study new observables, such as non-gaussianities, and to make more
accurate predictions of the observables we already measure.

The gauge invariant curvature perturbation on uniform density
hypersurfaces, $\zeta$, is a key variable when computing predictions
from inflation \cite{Bardeen:1983qw,Salopek:1990jq,ABMR,BKMR}. Its correlation
functions encode the relevant information about the micro-physics of
inflation and are also related with observables we can probe, such as
the temperature anisotropies in the CMB and the distribution of
galaxies in the sky. It is, therefore, fundamental to have a precise
knowledge about the evolution of this curvature perturbation. It is
well known that, on super-horizon scales and for adiabatic
perturbations, $\zeta$ is conserved at first order (see
e.g.~Ref.~\cite{WMLL}). This result has been extended to higher orders
using non-perturbative methods such as the $\delta N$
formalism \cite{LMS} and using other approaches (see
e.g.~Refs.~\cite{Rigopoulos:2003ak,LV}).  At second order in
cosmological perturbation theory, the conservation of the curvature
perturbation has been established for scalar perturbations in
Ref.~\cite{MW2004}.

However, the evolution equation at all scales and at second order in
the perturbative expansion including scalar, vector and tensor
perturbations has so far not been fully explored\footnote{Note, however, the early works of Tomita in Refs. \cite{Tomita1,Tomita2,Tomita3} regarding the interaction between scalar, vector and tensor modes in the Einstein-de Sitter universe, as well as more recent work in Ref. \cite{MMB} and references therein.}. Furthermore, at
second order, there exist different definitions of this gauge
invariant perturbation, and it is not clear that the conservation of
one implies the conservation of the others.

In this work, we aim to address both of these issues. We start by
reviewing the different versions of the gauge invariant curvature
perturbation on uniform density hypersurfaces and show how they are
related. We then derive the evolution equation for each convention and
compare the results. Besides the scalar contributions we also keep all
vector and tensor contributions, as well as the anisotropic
stress. Finally, we take the large scale limit and check for the
conditions of existence of conserved quantities.\\

The paper is structured as follows. In the next section, we present
the different conventions for the metric perturbations and give the
necessary gauge transformations. The different definitions of $\zeta$
are given in Section \ref{GI}, along with a number of auxiliary gauge
invariant quantities. A derivation of the evolution of $\zeta_2$ is
presented in Section \ref{evolution}. We then present our conclusions
in Section \ref{conclusion}.

\section{Definitions and Conventions}
\label{Conv}

In this initial section, we introduce our conventions for the metric
and stress-energy tensors. We simultaneously perform the 3+1
decomposition of space-time and and also the scalar-vector-tensor
(SVT) decomposition, in which tensors in the spatial slicing are split
into their scalar, divergence-free vector and transverse and traceless
tensor parts according to their transformation behaviour on the
3-dimensional submanifolds. For the most part, we follow the notation
of \cite{MW2008}. All quantities $T$ are expanded as
\bea
T=T_{0}+\delta T_1+\frac12\delta T_2+...\,,
\eea
in which the subscript denotes the order in perturbation theory. We will drop the subscript for the background quantities as the
unperturbed ones only appear at their definition.

Throughout the paper we assume a FLRW background spacetime with zero
spatial curvature, and use conformal time. Greek indices,
$\mu,\nu,\lambda$, range, from $0,\ldots 3$, while lower case Latin
indices, $i,j,k$, denote spatial indices ranging from $1,\ldots3$.

\subsection{Metric Tensor}

The metric tensor can be split in several different ways. We will show
four distinct ones, which vary in the way the spatial part of the
metric is arranged. The version that we will use in most of the
calculations below takes the following form, at all orders
\begin{align}
g_{00}=&-a^2\lb 1+2\phi\rb\,,\\
g_{i0}=&a^2\lb B_{,i}-S_i\rb\,,\\
g_{ij}=&a^2\lbs\delta_{ij}+2C_{ij}\rbs\,,
\end{align}
in which $\phi$ is the perturbation to the lapse, $B$ and $S_i$ are,
respectively, the scalar and vector parts of the shift and $C_{ij}$ is
the perturbation to the spatial part of the metric. The first
convention we will treat is defined by arranging $C_{ij}$
as\footnote{Note that the definition of the tensor perturbation is
slightly different from the one in Ref.~\cite{MW2008},
i.e.~$h_{ij}^{\rm{(MW)}}=2h_{ij}$.}
\begin{equation}
\label{gij}
C_{ij}=-\psi\delta_{ij}+E_{,ij}+F_{(i,j)}+h_{ij}\,,
\end{equation}
in which $\psi$ is the curvature perturbation in this metric
convention (as used by Mukhanov, Feldman and Brandenberger in Ref. \cite{MFB} and Malik and Wands in Ref. \cite{MW2008}, for example), $E$ and $F_i$ are, respectively, a scalar and a vector
part of the spatial metric and $h_{ij}$ is the tensor potential,
representing gravitational waves. The vector and tensor quantities
above obey
\bea
S^i_{\ ,i}=0\,,\ \ \ F^i_{\ ,i}=0\,,\ \ \ h^{ij}_{\ \ ,j}=0\,,
\ \ \ h^{i}_{\ i}=0\,.
\eea

This first convention for $\psi$ can be understood, at first order, as the perturbation to the intrinsic curvature, as explained in Appendix \ref{app3}. The other conventions do not have this property, but can be generally understood as perturbations to the scale factor $a(t)$. Appendix \ref{app3} also contains a definition of the scale factor from the extrinsic curvature, which is more easily relatable to the versions of $\psi$ given below.

A variation from the form \eqref{gij} consists of collecting the trace
of $C_{ij}$ in a single variable, here denoted by $\psi_T$. This split was used, e.g., by Bardeen in Ref. \cite{Bardeen} and also by Kodama and Sasaki in Ref. \cite{KodamaSasaki}, where $\psi_T$ was denoted by $H_L$. The
perturbation to the spatial part of the metric becomes
\be
\label{gijT}
C_{ij}=-\psi_T\delta_{ij}+E_{,ij}-\frac13\delta_{ij}\n^2E+F_{(i,j)}+h_{ij}\,,
\ee
which, upon comparison with the previous convention, \eqref{gij},
shows that the new curvature perturbation $\psi_T$ is given by, at all
orders
\be
\label{psT}
\psi_T=\psi-\frac13\n^2E\,.
\ee

The third kind of decomposition of $g_{ij}$ we will treat is similar
to the second one, Eq. \eqref{gijT}, but factors out the determinant
of the spatial part of the metric, instead of the trace. This is the decomposition used by Salopek and Bond in Ref. \cite{Salopek:1990jq} and also by Maldacena in Ref. \cite{Maldacena}. It can be
written as
\be
\label{gijD}
g_{ij}=a^2e^{2\psi_D}[e^{\omega}]_{ij}\,,
\ee
in which $\omega$ is a traceless tensor and $\psi_D$ is the curvature
perturbation of interest in this convention, defined by
$e^{6\psi_D}\equiv \det (g_{ij}/a^2)$. 
This quantity is usually interpreted as being a perturbation to the number of e-folds \cite{Salopek:1990jq}, $N$, given by $N=\ln a-\psi_D-\psi_D^2$. A related interpretation would be to think of it as a perturbation to the volume of spatial hypersurfaces, as it is proportional to the determinant of the spatial metric.
It can be shown \cite{MW2008}, that
the first and second order parts of $\psi_D$ are related by the
following expressions, in the notation of this work
\begin{align}
\psi_{D1}=&-\psi_{T1}=-\psi_1+\frac13\n^2E_1\,,\\
\psi_{D2}=&-\psi_{T2}-\frac23C_{1ij}C_1^{ij}=\\
=&-\psi_2+\frac13\n^2E_2-2\psi_1^2-\frac23h_{1ij}h_1^{ij}
+\frac43\psi_1\n^2E_1\nonumber\\
&-\frac43 h_1^{ij} \lb E_{1,ij}+F_{1i,j}\rb-\frac23 F_{1(i,j)}F_1^{j,i}-\frac23E_{1,ij}\lb2 F_1^{i,j}+E_1^{,ij}\rb\,.
\end{align}

The fourth convention is not a variation of $g_{ij}$ per se, but only a different way of defining the curvature perturbation. As with the third definition, Eq. \eqref{gijD}, we factor out the determinant of the spatial part of the metric, but in this case, we use the inverse metric to do so. Therefore, it is now defined as
\be
\label{gijI}
g^{ij}=a^{-2}e^{-2\psi_{I}}[e^{\omega_I}]^{ij},
\ee
in which, again, $\omega_I$ is a traceless tensor and $\psi_I$ is the new version of the curvature perturbation, determined by $e^{-6\psi_I}\equiv \det (g^{ij} a^2)$.
To our knowledge, this is the first time this definition has been used in the literature. Concerning its interpretation, it can still be seen as a perturbation to the scale factor and we find it to be equal to the integrated expansion, when the latter is evaluated in a comoving threading (see Appendix \ref{app3} for more details).
Comparing this new version of $\psi$ to the original one, we find the following relations
\begin{align}
\psi_{I1}=&-\psi_1+\frac13\n^2E_1,\\
\psi_{I2}=&-\psi_2+\frac13\n^2E_2-2\psi_1^2-\frac23h_{1ij}h_1^{ij}+\frac43\psi_1\n^2E_1+\frac13\lb B_{1,i}^{\vphantom{,i}}-S_{1i}\rb\lb B_1^{,i}-S_1^{i}\rb\nonumber\\
&-\frac43 h_1^{ij} \lb E_{1,ij}+F_{1i,j}\rb-\frac23 F_{1(i,j)}F_1^{j,i}-\frac23E_{1,ij}\lb2 F_1^{i,j}+E_1^{,ij}\rb.
\end{align}

We will use these four conventions to define different versions of
the gauge invariant curvature perturbation in the next section.

\subsection{Stress-energy Tensor}

As for the stress-energy tensor, it is defined in the so-called energy
frame \cite{EMM,HwangNoh}\footnote{This frame is defined by the
condition that the 4-velocity, $u^\mu$, is an eigenvector of
$T_{\mu\nu}$, with eigenvalue $\rho$. This is equivalent to setting
the energy flux, $q^\mu$, to zero.}, including anisotropic stress:
\be
T_{\mu\nu}=\lb P+\rho\rb u_\mu u_\nu+P g_{\mu\nu}+\pi_{\mu\nu}\,,
\ee
with
\be
\label{constrPi}
\pi^\mu_{\ \mu}=0\,,\ \  \text{  and }\pi_{\mu\nu}u^\mu=0\,. 
\ee
The perturbative expansion is the standard one, as given
in \cite{MW2008}. We merely reproduce the SVT splitting of
$\pi_{2\mu\nu}$, as it is non-trivial at this order:
\bea
\pi_{200}=0,\ \ \ \pi_{2i0}=-2\pi_{1ij}\lb v_{V1}^j+v_{1}^{,j}\rb\,,\nonumber\\
\pi_{2ij}=a^2\lbs\Pi_{2ij}+\Pi_{2(i,j)}+\Pi_{2,ij}-\frac{1}{3}\delta_{ij}\n^2\Pi_2\rbs+\frac43\delta_{ij}\pi_{1kl}C_1^{kl}\,,
\eea
where the quantities $\Pi$, $\Pi_i$ and $\Pi_{ij}$ are, respectivelly, the scalar, vector and tensor parts of the linear piece of the anisotropic stress tensor, $v$ and $v_V^i$ are the scalar and vector components of the velocity perturbation and $C_{1 ij}$ is the spatial metric perturbation, as given in equation~\eqref{gij}. The
terms that arise in $\pi_{2i0}$ and $\pi_{2ij}$ which depend on the first order anisotropic stress, $\pi_{1ij}$, are essential in order
to not violate the constraints that define the anisotropic
stress, Eqs.~\eqref{constrPi}. Should these terms not be included, second
order contributions of the anisotropic stress to the evolution equations are not well taken into
account. That seems to be the case for the energy conservation
equation given in \cite{MW2008}, which differs from our result presented
below in equation \eqref{evorho2}.

\subsection{The gauge problem}

Most of the quantities defined above will depend on a choice of
coordinates on the perturbed manifold, the so-called gauge
choice \cite{Bardeen,KodamaSasaki,MFB,MM2008,MW2008}. A gauge generator
vector field $\xi$ is defined to parametrise the change of coordinate
system. The gauge transformation it induces in a tensor $T$ is given
by
\bea
\widetilde T=e^{\L_\xi}T\,,
\eea
with $\L_\xi$ being the Lie derivative in the direction of $\xi$. The
transformation rules for the perturbations are obtained by expanding
the relation above order by order. Up to second order, the gauge
transformations are given by
\bea
\widetilde{\delta T_1}=\delta T_1+\L_{\xi_1}T_0\,,\\
\widetilde{\delta T_2}=\delta T_2+\L_{\xi_2}T_0+\L_{\xi_1}^2T_0
+2\L_{\xi_1}\delta T_1\,,\label{gauge2}
\eea
where the gauge generator was also expanded order by order as
$\xi^\mu=\xi_1^\mu+\frac12 \xi_2^\mu+\dots$. It can be 
decomposed further into scalar and vector parts as
\be
\lb\xi^\mu\rb=\lb\alpha ,\beta^{,i}+\gamma^i\rb\,.
\ee
Applying these transformations to the metric tensor, one finds the
first order metric potentials of the original convention,
Eq.~\eqref{gij}, to change to
\begin{align}
&\wt{\phi}_1= \phi_1+\H\alpha_1+\alpha_1'\,,\ \ \ \ \ 
\wt{\psi}_1=\psi_1-\H\alpha_1\,,\\
&\wt{E}_1= E_1+\beta_1\,,\ \ \ \ \ \ \ \quad\quad\>\wt{B}_1=B_1-\alpha_1+\beta_1'\,,\\
&\wt{F}_1^i= F_1^i+\gamma_1^i\,,\ \ \ \ \ \ \ \quad\quad\ \wt{S}_1^i=S_1^i-\gamma_1^{i\prime}\,,\\
&\wt{h}_1^{ij}= h_1^{ij}\,,
\end{align}
and the first order fluid quantities to transform to
\begin{align}
&\wt{\delta\rho_1}=\delta\rho_1+\alpha_1\rho'\,,\ \ \ \ \ \wt{\delta P_1}=\delta P_1+\alpha_1P'\,,\\
&\wt{v_1}=v_1-\beta_1'\,,\ \ \ \ \ \quad\quad\wt{v_{V1}^i}=v_{V1}^i-\gamma_{1}^{i\prime}\,.
\end{align}
The gauge transformations at second order are given in Appendix \ref{app1}.

The ambiguities created by this gauge issue are solved by ``fixing'' or
choosing a gauge and hence constructing gauge invariant
quantities \cite{Bardeen,KodamaSasaki,MFB,MW2004,MW2008}. 
In the following section we derive gauge invariant quantities,
specifying a particular gauge by choosing convenient temporal and
spatial hypersurfaces, or specifying the slicing and threading,
respectively (for a non-technical introduction to the gauge issue in
cosmology see Ref.~\cite{MalikMatravers}).
The gauge invariant quantities are uniquely defined by specifying the
hypersurface (at all orders), and can therefore also be evaluated in
other gauges, if necessary.

\section{Gauge invariant quantities}
\label{GI}

The method we use to generate gauge invariant variables starts with
performing a gauge transformation on a variable of interest,
e.g.~$\psi_1$. One then substitutes the gauge generator components
$\xi_1^\mu$ with those obtained by solving a gauge fixing constraint,
e.g.~$\widetilde{\dr_1}=0$. The end result is a gauge invariant
quantity, e.g.~the curvature perturbation in uniform density
hypersurfaces, $\zeta_1$. We apply this method for the quantities of
interest in the subsections below (for details see
e.g.~Refs.~\cite{MW2008,MM2008}).

\subsection{Curvature perturbation on uniform density hypersurfaces}

The focus of this work is the curvature perturbation on uniform
density hypersurfaces $\zeta$. As was already sketched above, it is
defined to be equal to $-\psi$ in the gauge in which the density field
is uniform ($\widetilde{\dr}=0$). Starting with our first convention
for the metric, Eq.~\eqref{gij}, this condition is sufficient to fully
construct $\zeta$ at first order as (\cite{Bardeen:1983qw,WMLL})
\be
\label{z1}
\zeta_1\equiv-\psi_1-\H\frac{\dr_1}{\rho'}\,.
\ee
However, at second order, one is also forced to specify the first
order gauge to define this curvature perturbation unambiguously. For
this convention of the metric tensor, Eq.~\eqref{gij}, we will use the
following gauge conditions to define $\zeta_2$ (\cite{MW2004})
\be
\zeta_2\equiv-\widetilde{\psi_2}~, \text{ if }\widetilde{\dr_2}=\widetilde{\dr_1}=\widetilde{E_1}=0~,~~\widetilde{F_{1}^i}=0\,.
\ee
These add a flat threading to the uniform density gauge. The general
formal expression for $\zeta_2$ is given in \cite{MW2008}. In full detail,
the formula is rather complicated and we write it here with the
r.h.s.~evaluated in flat gauge\footnote{Flat gauge is defined by the
conditions $\psi=E=F^i=0$.},
\begin{align}
\label{z2}
\zeta_2=&-\frac{\H}{\rho'}\dr_2+\frac{1}{\rho'^2}\lb 2\H^2+\H'-\H\frac{\rho''}{\rho'}\rb\dr_1^2+\frac{2\H}{\rho'^2}\dr_1\dr'_1-\frac 1 {2 \rho'^2} \dr_{1,k}\dr_1^{,k}-\frac 1 {\rho'}\lb B_{1,k}-S_{1k}\rb\dr_1^{,k}\nonumber\\
&+\nabla^{-2}\lbc\frac12\lbs\frac 1{\rho'^2}\dr_1^{,i}\dr_1^{,j}+\frac 2{\rho'}\lb\dr_1^{,(i} B_1^{,j)}-\dr_1^{,(i}S_1^{j)}\rb\rbs_{,ij}+\frac{1}{\rho'}\lb h_1^{ij\prime}+2\H h_1^{ij}\rb \dr_{1,ij}\rbc\,.
\end{align}
We can see that, in contrast with the first order result, the second
order $\zeta$ is much harder to relate to density perturbations in
flat gauge, given the presence of vectors and tensors. In spite of
this, this expression is still useful in writing the gauge invariant
curvature perturbation in terms of multiple scalar fields, as is done
in \cite{CNM,DEFMS}.
\\

Let us now move to the second convention of the metric,
Eq.~\eqref{gijT}. In this case, $\widetilde{\dr}=0$ is no longer a
sufficient gauge condition to define an invariant, even at first
order; one must also specify the scalar part of the threading, due to
the inclusion of $E$ in the definition of $\psi_T$ (see
Eq.~\eqref{psT}). The extra condition that is most often chosen is
$\widetilde{v_1}=0$, which results in the following
expression\footnote{An alternative choice would be
$\widetilde{E_1}=0$, but that would simply result in the expression
for the original metric convention, as
$\widetilde{\psi_{T1}}=\widetilde{\psi_1}$.}
\be
\label{zT1w}
\zeta_{T1}=-\psi_{T1}-\frac{\H}{\rho'}\dr_1+\frac{1}{3}\n^2\int v_1 d\tau\,,
\ee
in which the integral in conformal time is indefinite. The
introduction of these integrals is the disadvantage of using the gauge
condition, $\widetilde{v_1}=0$. This might be problematic, as this
condition only sets the gauge up to an arbitrary function of the
spatial coordinates, which, in turn, might spoil the gauge invariance of
the new variable. In spite of this, it is possible to construct a gauge
invariant quantity, by defining it to be
\be
\label{zT1}
\zeta_{T1}\equiv\zeta_1+\frac{1}{3}\n^2\int \J_1 d\tau\,.
\ee
with $J_1$ being the gauge invariant velocity on flat hypersurfaces, defined by
\be
\label{j1}
\J_1=E_1'+v_1\,.
\ee
While the integral in Eq.~\eqref{zT1} is still indefinite, the integrand is gauge invariant and, therefore, this is the
definition we use.

At second order, one sets the second order gauge in the same way,
i.e.~$\wt{\dr_2}=\widetilde{v_2}=0$ and, to avoid additional issues
with indefinite integrals, one can choose
$\widetilde{\dr_1}=\widetilde{E_1}=\widetilde{F_1^i}=0$ for the first
order gauge fixing. With this choice, we find
\be
\label{zT2}
\zeta_{T2}=\zeta_2+\frac{1}{3}\n^2\int \J_2 d\tau,
\ee
in which $\J_2$ is the second order equivalent of $\J_1$ in this
gauge, i.e.~it equals $E_2'+v_2$ in the gauge obeying
$\widetilde{\dr_1}=\widetilde{E_1}=\widetilde{F_1^i}=0$. As is visible
in the expression above, Eq.~\eqref{zT1}, the only variable of
interest is $\n^2\J_2$ and hence, for shortness of presentation, that
is all we show below, with the r.h.s.~evaluated in flat gauge
\begin{align}
\label{j2}
\n^2\J_2=&\n^2v_2+\frac{2}{\rho'}\lbs\dr_1\lb\H(v_{V1}^i+v_1^{,i})-v_{V1}^{i\prime}-v_1'^{,i}\rb\rbs_{,i}+\lbs\frac{\dr_{1,i}\dr_{1}^{,i}}{2\rho'^2}+\frac{\lb B_1^{,i}-S_1^i\rb \dr_{1,i}}{\rho'}\right.\\
&\left.+\nabla^{-2}\lbc-\frac32\lbs\frac 1{\rho'^2}\dr_1^{,i}\dr_1^{,j}+\frac 2{\rho'}\lb\dr_1^{,(i} B_1^{,j)}-\dr_1^{,(i}S_1^{j)}\rb\rbs_{,ij}-\frac{3}{\rho'}\lb h_1^{ij\prime}+2\H h_1^{ij}\rb \dr_{1,ij}\rbc\rbs'\,.\nonumber
\end{align}
As we will see in Section \ref{evolution}, this quantity is
relevant regardless of the choice of convention for the metric, as it
will appear in the evolution equation for the curvature perturbation.
\\

Let us now turn to the third convention of the metric,
Eq.~\eqref{gijD}. For this case, $\zeta_D$ will be defined as being
equal to $\psi_D$ instead of $-\psi_D$, in order to keep the same sign
as $\zeta$. Starting at first order, we see that we get either
$\zeta_{D1}=\zeta_1$ or $\zeta_{D1}^{(v)}=\zeta_{T1}$, depending on
whether we choose $\widetilde{E_1}=0$ or $\widetilde{v_1}=0$,
respectively, for fixing the threading. The second order result is
more interesting, as there is no gauge fixing for which it is equal to
either of the other definitions above. In the most conservative case,
the choice of gauge fixing is $\widetilde{\dr_2}=\widetilde{E_2}=0$ at
second order and
$\widetilde{\dr_1}=\widetilde{E_1}=\widetilde{F_1^i}=0$ at first
order. This results in\footnote{This result is well known in the case without tensors. See, for example, Refs. \cite{BKMR,DEFMS,LV}.}
\be
\label{zD2}
\zeta_{D2}=\zeta_2-\frac23h_{1ij}h_1^{ij}-2\zeta_1^2\,.
\ee
A different gauge fixing is $\widetilde{\dr_2}=\widetilde{v_2}=0$ and
$\widetilde{\dr_1}=\widetilde{v_1}=\widetilde{v_{V1}}^i=0$, for which
the result is
\begin{align}
\label{zD2v}
\zeta_{D2}^{(v)}=&\zeta_2+\frac13 \int \n^2\J_2d\tau-\frac23h_{1ij}h_1^{ij}-2\zeta_1^2\\
&+2\zeta_{1,i}\int\lb \J_{1}^{^,i}+\V_1^{i}\rb d\tau+\frac23\int\lbc\lbs\lb \J_{1}^{^,i}+\V_1^{i}\rb\U_1\rbs_{,i} +\n^2\J_{1,i}\int \lb \J_{1}^{^,i}+\V_1^{i}\rb d\tau'\rbc d\tau\,,\nonumber
\end{align}
in which $\V_1^i$ is the gauge invariant velocity vector perturbation
in flat hypersurfaces and $\U$ is the gauge invariant lapse
perturbation in uniform density hypersurfaces. In a general gauge,
these quantities are given by
\begin{align}
&\V_1^i=v_{V1}^i+F_1^{i\prime},\\
&\U_1=\phi_1-\H\frac{\dr_1}{\rho'}-\lb\frac{\dr_1}{\rho'}\rb'\,.
\end{align}

For the forth version of the curvature perturbation, Eq. \eqref{gijI}, the procedure is very similar to the one for the third convention. As in the previous case, the first order quantities obey $\zeta_{I1}=\zeta_1$ or $\zeta_{I1}^{(v)}=\zeta_{T1}$, depending on whether $\widetilde{E_1}=0$ or $\widetilde{v_1}=0$ is chosen for setting the threading. At second order, the results are
\be
\label{zI2}
\zeta_{I2}=\zeta_2-\frac23h_{1ij}h_1^{ij}-2\zeta_1^2+\frac13\lb \W_{1i}-\V_{1i}+\A_{1,i}-\J_{1,i}\vphantom{\A_{1}^{,i}}\rb\lb \W_{1}^i-\V_{1}^i+\A_{1}^{,i}-\J_{1}^{,i}\rb\,,
\ee
if the gauge is fixed with $\widetilde{\dr_2}=\widetilde{E_2}=0$ and $\widetilde{\dr_1}=\widetilde{E_1}=\widetilde{F_1^i}=0$, and 
\begin{align}
\label{zI2v}
\zeta_{I2}^{(v)}=&\zeta_2+\frac13 \int \n^2\J_2d\tau-\frac23h_{1ij}h_1^{ij}-2\zeta_1^2+\frac13\lb \W_{1i}+\A_{1,i}\vphantom{\A_{1}^{,i}}\rb\lb \W_{1}^i+\A_{1}^{,i}\rb\\
&+2\zeta_{1,i}\int\lb \J_{1}^{^,i}+\V_1^{i}\rb d\tau+\frac23\int\lbc\lbs\lb \J_{1}^{^,i}+\V_1^{i}\rb\U_1\rbs_{,i} +\n^2\J_{1,i}\int \lb \J_{1}^{^,i}+\V_1^{i}\rb d\tau'\rbc d\tau\,,\nonumber
\end{align}
when the gauge choice is  $\widetilde{\dr_2}=\widetilde{v_2}=0$ and $\widetilde{\dr_1}=\widetilde{v_1}=\widetilde{v_{V1}}^i=0$. The new first order gauge invariant quantities that appear are the vector velocity in zero shift gauge, $\W_1^i$, and the
momentum perturbation in uniform density gauge, $\A_1$. They are given
by
\begin{align}
&\W_1^i=v_{V1}^i-S_1^i,\\
&\A_1=v_1+B_1+\frac{\dr_1}{\rho'}.
\end{align}

\subsection{Non-adiabatic pressure}

One of the quantities determining the evolution of the curvature
perturbation is the non-adiabatic
pressure \cite{Mollerach:1989hu,WMLL,nonad,KodamaSasaki,Bardeen}. It is defined as the
deviation from the adiabatic relation as
\bea
\label{nad}
\delta P=c_\text{s}^2\delta\rho+\delta P_\text{nad}\,,
\eea
with $c_{\rm s}$ the adiabatic sound speed defined as $c_{\rm
s}^2=P'/\rho'$. At first order, this definition automatically
generates a gauge invariant quantity, but, at second order, this is
not sufficient and many choices can be made. Our first choice is to
define it as the gauge invariant quantity that reduces to
Eq.~\eqref{nad} in the gauge in which
$\widetilde{\dr_1}=\widetilde{E_1}=\widetilde{F_{1}^i}=0$. In a
general gauge, this quantity is given by
\begin{align}
\label{nad2}
\delta P_{\text{nad}\, 2}=&\delta P_2-c_{\rm s}^2\dr_2-\frac{2}{\rho'}\dr_1\delta P_1'+\lb\frac{P''}{\rho'^2}-\frac{P'\rho''}{\rho'^3}\rb\dr_1^2\nonumber\\
&+\frac{2c_s^2}{\rho'}\dr_1\dr_1'-2\lb F_1^i+E_1^{,i}\rb\delta P_{\text{nad}\, 1,i}\,.
\end{align}
With the different choice of threading,
$\widetilde{v_1}=\widetilde{v_{V1}}^i=0$, one finds instead
\begin{align}
\label{nad2v}
\delta P_{\text{nad}\, 2}^{(v)}=\delta P_{\text{nad}\, 2}+2\delta P_{\text{nad}\, 1,i}\int\lb \V_{1}^i+\J_1^{,i}\rb d\tau\,.
\end{align}
For a barotropic fluid, with $P=P(\rho)$, both expressions vanish, as
can be easily checked by evaluating them in their defining gauge,
i.e.~with $\dr_1=E_1=F_{1}^i=0$.

The quantities presented so far include the full set of gauge invariant quantities required for the full derivation of the evolution equations below.

\section{Evolution equations}\label{evolution}

In this section, we present the derivation of the evolution equations
for all versions of $\zeta$. Our strategy consists of calculating the
derivative of expression \eqref{z2} and using only the perturbed
energy-momentum conservation equations up to second order to simplify
the result. Lastly, we substitute the gauge dependent variables for
gauge invariant ones, using the expressions found in the previous
Section, to arrive at our final result. Having found the result for
$\zeta_2$ in the original convention of the metric, Eq.~\eqref{gij},
we then rewrite the evolution equation in terms of the different
definitions of $\zeta$.

\subsection{Fluid equations}

Energy-momentum conservation, $T^{\mu\nu}_{\ \ ;\nu}=0$, governs the
evolution of the fluid density and velocity. For simplicity, we
present these evolution equations in flat gauge. The first order
energy conservation equation is given by
\be
\label{evorho1}
\dr_1'+3\H\lb\dr_1+\delta P_1\rb+\lb\rho+P\rb\n^2v_1=0~\,,
\ee
while momentum conservation is
\be
\label{evov1}
\delta P_{1,k}+(\rho+P)\lbs Z_{1k}'+\phi_{1,k}+\lb1
-3 c_s^2\rb\H Z_{1k}\rbs+\frac23 \n^2\Pi_{1,k}+\frac12 \n^2\Pi_{1k}=0\,,
\ee
where the momentum perturbation $Z_1^{\ k}$ is given by
\be
Z_1^{\ k}=v_{V1}^{\ \ k}-S_1^{\ k}+B_1^{\ ,k}+v_1^{\ ,k}\,.
\ee
At second order, we only require the energy conservation equation, which is
\begin{align}
\label{evorho2}
\delta\rho'_2=&-3\H\lb\dr_2+\delta P_2\rb-\lb\rho+P\rb\n^2v_2-2\lb\delta P_1+\dr_1\rb \n^2 v_1-2\dr_{1,k}\lb v_{V1}^{k}+v_1^{,k}\rb\nonumber\\
&-2 \delta P_{1,k} Z_1^{\ k}-\lb \rho+P\rb \lbs\vphantom{\lb v_{V1}^{\ \ k}+v_1^{\ ,k}\rb} 4 Z'_{1k} Z_1^{\ k}+2\lb 1-3 c_s^2\rb \H Z_{1k}Z_1^{\ k} + 2\phi_{1,k}Z_1^{\ k}\right.\nonumber\\
&\left.+2\phi_{1,k}\lb v_{V1}^{\ \ k}+v_1^{\ ,k}\rb+2\phi_1 \n^2 v_1-4h'_{1ij}h_1^{\ ij}\rbs-Z_1^{\ k}\lb\frac43 \n^2\Pi_{1,k}+\n^2\Pi_{1k}\rb\\
&-2\lb h'_{1ij}+v_{V1i,j}+v_{1,ij}\rb\lb\Pi_1^{\ ij}+\Pi_1^{\ (i,j)}+\Pi_1^{\ ,ij}-\frac13\delta^{ij}\n^2\Pi_1\rb\,.\nonumber
\end{align}
The above equations are sufficient to derive evolution equations
for the curvature perturbation at first and at second order \cite{WMLL}.

\subsection{Evolution of the curvature perturbation}\label{EvoCurv}

We can now derive the evolution equation for the curvature
perturbation on uniform density hypersurfaces. We follow the strategy
stated at the beginning of this section. At first order, the result is
well known to be
\be
\label{z1evo}
\zeta_1'=-\frac13\n^2\J_1-\H\frac{\delta P_{\text{nad}\, 1}}{\rho+P}~,
\ee
where only the first order energy conservation equation was used. On
large scales (``$\n\rightarrow0$") and in the absence of non-adiabatic
pressure, one finds the familiar conservation equation $\zeta_1'=0$.

For the other conventions for the curvature perturbation, $\zeta_{T1}$, $\zeta_{D1}^{(v)}$ and $\zeta_{I1}^{(v)}$, the evolution equation at first order is the
same and is given by
\be
\label{zetaT1p}
\zeta_{T1}'=-\H\frac{\delta P_{\text{nad}\, 1}}{\rho+P}~,
\ee
which shows these versions of $\zeta_1$ are conserved at all scales,
when non-adiabatic pressure is negligible \cite{LMS,LM}.

At second order, the complexity increases. The procedure to obtain the
final result is as follows: use the energy conservation equation at
first (Eq.~\eqref{evorho1}) and second order (Eq.~\eqref{evorho2}) to
substitute for $\dr_1'$ and $\dr_2'$ and substitute
$\lb\frac43 \n^2\Pi_{1,k}+\n^2\Pi_{1k}\rb$ with the momentum
conservation equation, Eq.~\eqref{evov1}. The last step is to use the
defining expressions of the gauge invariants to eliminate all gauge
dependant variables. The final result is given by\footnote{Note the
absence of inverse Laplacians. That is explained by an exact
cancellation between the terms in $\zeta_2'$ and those in $\n^2J_2$,
as can be shown by comparing equations \eqref{j2} and \eqref{z2}.}
\begin{align}
&\lb-\zeta_2+2\zeta_1^2-\frac13\lb \W_{1i}+\A_{1,i}\vphantom{\A_{1}^{,i}}\rb\lb \W_{1}^i+\A_{1}^{,i}\rb+\frac23h_{1ij}h_1^{ij}\rb'=\\
&\frac13\n^2\J_2+\H\frac{\delta P_{\text{nad}\, 2}}{\rho+P}-2\H\frac{\delta P_{\text{nad}\, 1}^2}{(\rho+P)^2}+\frac{2}{3}\lbs\U\lb \V_1^i+\J_1^{,i}\rb\rbs_{,i}+2\zeta_{1,i}\lb \V_1^i+\J_1^{,i}\rb\nonumber\\
&-\frac{2\H}{\rho'}\lb\Pi_{1ij}+\Pi_{1(i,j)}+\Pi_{1,ij}-\frac{1}{3}\delta_{ij}\n^2\Pi_1\rb\lb h_1^{ij\prime}+\V_1^{i,j}+\J_1^{,ij}\rb\nonumber.
\end{align}
We are now able to identify the different terms that source the evolution of $\zeta_2$. We note, in particular, the appearance of vector and tensor source terms as well as the anisotropic stress which did not appear at first order in this equation\footnote{Note however, that the scalar part of the anisotropic stress tensor would source the evolution of $\zeta$ at first order by acting on the evolution of $\n^2J$. This can be seen more clearly by deriving Eq.~\eqref{z1evo} and using the momentum conservation equation, Eq.~\eqref{evov1}, to substitute for $\n^2J$:
\be
\zeta_1''+\H\zeta_1'-\frac{P'}{3(\rho+P)}\n^2A_1-\frac13\n^2\Phi_1+\left(\H\frac{\delta P_{\text{nad}\, 1}}{\rho+P}\right)'+\H^2\frac{\delta P_{\text{nad}\, 1}} {\rho+P}-\frac{\n^2\delta P_{\text{nad}\, 1}}{3 (\rho+P)}-\frac{2}{9 (\rho+P)}\n^2\n^2\Pi_1=0\,,
\ee
in which $\Phi_1$ is one of the Bardeen potentials, given in terms of our variables as $\Phi_1=\Upsilon_1+\H(A_1-J_1)+(A_1-J_1)'$.
}.

We are now in the position to substitute for the other versions of
$\zeta$ and find their evolution equations. For $\zeta_{2T}$, we find
\begin{align}
&\lb-\zeta_{2T}+2\zeta_1^2-\frac13\lb \W_{1i}+\A_{1,i}\vphantom{\A_{1}^{,i}}\rb\lb \W_{1}^i+\A_{1}^{,i}\rb+\frac23h_{1ij}h_1^{ij}\rb'=\\
&\H\frac{\delta P_{\text{nad}\, 2}}{\rho+P}-2\H\frac{\delta P_{\text{nad}\, 1}^2}{(\rho+P)^2}+\frac{2}{3}\lbs\U\lb \V_1^i+\J_1^{,i}\rb\rbs_{,i}+2\zeta_{1,i}\lb \V_1^i+\J_1^{,i}\rb\nonumber\\
&-\frac{2\H}{\rho'}\lb\Pi_{1ij}+\Pi_{1(i,j)}+\Pi_{1,ij}-\frac{1}{3}\delta_{ij}\n^2\Pi_1\rb\lb h_1^{ij\prime}+\V_1^{i,j}+\J_1^{,ij}\rb\nonumber,
\end{align}
while $\zeta_{D2}$ evolves as
\begin{align}
&\lb-\zeta_{D2}-\frac13\lb \W_{1i}+\A_{1,i}\vphantom{\A_{1}^{,i}}\rb\lb \W_{1}^i+\A_{1}^{,i}\rb\rb'=\\
&\frac13\n^2\J_2+\H\frac{\delta P_{\text{nad}\, 2}}{\rho+P}-2\H\frac{\delta P_{\text{nad}\, 1}^2}{(\rho+P)^2}+\frac{2}{3}\lbs\U\lb \V_1^i+\J_1^{,i}\rb\rbs_{,i}+2\zeta_{D1,i}\lb \V_1^i+\J_1^{,i}\rb\nonumber\\
&-\frac{2\H}{\rho'}\lb\Pi_{1ij}+\Pi_{1(i,j)}+\Pi_{1,ij}-\frac{1}{3}\delta_{ij}\n^2\Pi_1\rb\lb h_1^{ij\prime}+\V_1^{i,j}+\J_1^{,ij}\rb\nonumber,
\end{align}
and the result for $\zeta_{D2}^{(v)}$ is
\begin{align}
&\lb-\zeta_{D2}^{(v)}-\frac13\lb \W_{1i}+\A_{1,i}\vphantom{\A_{1}^{,i}}\rb\lb \W_{1}^i+\A_{1}^{,i}\rb\rb'=\H\frac{\delta P_{\text{nad}\, 2}^{(v)}}{\rho+P}-2\H\frac{\delta P_{\text{nad}\, 1}^2}{(\rho+P)^2}\\
&-\frac{2\H}{\rho'}\lb\Pi_{1ij}+\Pi_{1(i,j)}+\Pi_{1,ij}-\frac{1}{3}\delta_{ij}\n^2\Pi_1\rb\lb h_1^{ij\prime}+\V_1^{i,j}+\J_1^{,ij}\rb\nonumber.
\end{align}
The simplest evolutions equations are found for the $\zeta_{I2}$ and $\zeta_{I2}^{(v)}$  versions of the gauge invariant curvature perturbation. They are given by
\begin{align}
&\lb-\zeta_{I2}+\frac13\lb \V_{1i}+\J_{1,i}\vphantom{\A_{1}^{,i}}\rb\lb \V_{1}^i+\J_{1}^{,i}-2\W_{1}^i-2\A_{1}^{,i}\rb \rb'=\\
&\frac13\n^2\J_2+\H\frac{\delta P_{\text{nad}\, 2}}{\rho+P}-2\H\frac{\delta P_{\text{nad}\, 1}^2}{(\rho+P)^2}+\frac{2}{3}\lbs\U\lb \V_1^i+\J_1^{,i}\rb\rbs_{,i}+2\zeta_{I1,i}\lb \V_1^i+\J_1^{,i}\rb\nonumber\\
&-\frac{2\H}{\rho'}\lb\Pi_{1ij}+\Pi_{1(i,j)}+\Pi_{1,ij}-\frac{1}{3}\delta_{ij}\n^2\Pi_1\rb\lb h_1^{ij\prime}+\V_1^{i,j}+\J_1^{,ij}\rb,\nonumber
\end{align}
and 
\begin{align}
\label{zI2p}
&-\zeta_{I2}^{(v)}{}'=\H\frac{\delta P_{\text{nad}\, 2}^{(v)}}{\rho+P}-2\H\frac{\delta P_{\text{nad}\, 1}^2}{(\rho+P)^2}\\
&-\frac{2\H}{\rho'}\lb\Pi_{1ij}+\Pi_{1(i,j)}+\Pi_{1,ij}-\frac{1}{3}\delta_{ij}\n^2\Pi_1\rb\lb h_1^{ij\prime}+\V_1^{i,j}+\J_1^{,ij}\rb\nonumber.
\end{align}
This final expression, like its first order version, Eq. \eqref{zetaT1p}, shows that, in the absence of non-adiabatic pressure and anisotropic stress, this version of the curvature perturbation is conserved on all scales. While this is interesting, in order for this result to be useful, one would likely be forced to estimate the integrals in the defining expression for $\zeta_{I2}$, Eq. \eqref{zI2v}. This is not likely to be straightforward, given the indeterminate nature of the integrals. This evolution equation matches the results of Ref. \cite{Enqvist:2006fs} for the integrated expansion in the absence of anisotropic stress, obtained in the covariant approach.

\subsection{Large scale approximation}
Here we perform the large scale approximation, by neglecting all terms with spatial derivatives in the equations above \footnote{This is generally well motivated in the case of some metric potentials, as one expects the perturbed metric to approach the background metric on large scales \cite{LMS}, and we will assume the same is true for the matter variables, including the anisotropic stress. Should this assumption not hold for the particular model under study, then the results in this section are not valid and one should use the full results from section \ref{EvoCurv}.}.
This limit simplifies the evolution equations to
\be
\lb-\zeta_2+2\zeta_1^2-\frac13\W_{1i}\W_{1}^i+\frac23h_{1ij}h_1^{ij}\rb'=\H\frac{\delta P_{\text{nad}\, 2}}{\rho+P}-2\H\frac{\delta P_{\text{nad}\, 1}^2}{(\rho+P)^2}-\frac{2\H}{\rho'}\Pi_{1ij}h_1^{ij\prime}\,,
\ee
for $\zeta_2$, here representing both the original $\zeta_2$ and $\zeta_{T2}$, and 
\be
\lb-\zeta_{D2}-\frac13\W_{1i}\W_{1}^i\rb'=\H\frac{\delta P_{\text{nad}\, 2}}{\rho+P}-2\H\frac{\delta P_{\text{nad}\, 1}^2}{(\rho+P)^2}-\frac{2\H}{\rho'}\Pi_{1ij}h_1^{ij\prime}\,,
\ee
for the evolution of both $\zeta_{D2}$ and $\zeta_{D2}^{(v)}$, and
\begin{align}
\lb-\zeta_{I2}+\frac13\V_{1i}\lb\V_{1}^i-2\W_{1}^i\rb\rb'=\H\frac{\delta P_{\text{nad}\, 2}}{\rho+P}-2\H\frac{\delta P_{\text{nad}\, 1}^2}{(\rho+P)^2}-\frac{2\H}{\rho'}\Pi_{1ij}h_1^{ij\prime}\,,\\
-\zeta_{I2}^{(v)}{}'=\H\frac{\delta P_{\text{nad}\, 2}}{\rho+P}-2\H\frac{\delta P_{\text{nad}\, 1}^2}{(\rho+P)^2}-\frac{2\H}{\rho'}\Pi_{1ij}h_1^{ij\prime}\,.
\end{align}
Note that, in all cases above, the
pairs are equal in the large scale approximation, except for $\zeta_{I2}$ and $\zeta_{I2}^{(v)}$, which have a different contribution from vector perturbations. We give the expressions for each version in this approximation in Appendix \ref{app2}. From this result,
one can see that, even in the absence of the scalar non-adiabatic pressure, $\delta P_{\text{nad}}$,
neither curvature perturbation is conserved,
\be
\lb-\zeta_2+\frac23h_{1ij}h_1^{ij}-\frac13 W_1^iW_{1i}\rb'=\lb-\zeta_{D2}-\frac13 W_1^iW_{1i}\rb'=-\zeta_{I2}^{(v)}{}'
=-\frac{2\H}{\rho'}\Pi_{1ij}h_1^{ij\prime}\,.
\ee
However, if the traceless, transverse part of the anisotropic stress, $\Pi_{1ij}$, is negligible, $\zeta_{I2}^{(v)}$ is in fact conserved
\be
\zeta_{I2}^{(v)}{}'=\lb\zeta_{D2}+\frac13 W_1^iW_{1i}\rb'=\lb\zeta_2+\frac13 W_1^iW_{1i}-\frac23h_1^{ij}h_{1ij}\rb'=0\,.
\ee
Although $\zeta_{I2}^{(v)}$ is exactly conserved, the difference between $\zeta_{I2}^{(v)}{}'$ and $\zeta_{D2}'$ only depends on vector perturbations, which are usually negligible. Moreover, using the vector part of the momentum conservation equation, Eq. \eqref{evov1}, in the absence of anisotropic stress, we find the evolution of $W_1^i$ is given by
\be
W_{1i}^{\prime}+\H (1-3c_s^2) W_{1i}=0.
\ee
Thus, this vector perturbation is conserved during radiation domination ($c_s^2=1/3$) and, as a consequence, $\zeta_{D2}$ is exactly conserved during that epoch. In the general case, we may therefore write the evolution of $\zeta_{D2}$ on large scales as 
\be
\zeta_{D2}'=-\frac23\H (1-3c_s^2)W_1^iW_{1i}\,,
\ee
showing again that it may only have an appreciable evolution if the vector modes are large.

The evolution equations simplify further in Einstein gravity, as, in the absence of
anisotropic stress, tensor modes stop evolving and hence
this new conservation law converges fairly quickly to the conservation
of $\zeta_2$ itself.
Hence, for Einstein gravity, all versions of the curvature perturbation are conserved up to second order on
large scales, if both the non-adiabatic pressure and the
anisotropic stress are negligible. However, should the evolution of vectors and tensors be appreciable, the version of $\zeta$ which is conserved is $\zeta_{I}^{(v)}$, i.e., the version defined by the determinant of $g^{ij}$ and by using a comoving threading to fix the gauge.

\section{Conclusion}\label{conclusion}

We obtained the evolution equation for the curvature perturbation at
second order in cosmological perturbation theory, valid on all scales. With the inclusion of vectors, tensors and anisotropic stress,
this result allows for high precision calculations of correlation functions on all scales. We derive this for six different definitions of $\zeta$, based on several different splits of the spatial metric and on various choices of the defining gauge. The results for the evolution equations show a substantial difference in apparent complexity, being simpler when the threading defining $\zeta$ was chosen to be the comoving one, i.e. $\widetilde{v}^i=0$. Eq. \eqref{zI2p} for the evolution of $\zeta_{I2}^{(v)}$ is particularly short, but its usefulness is unclear due to the existence of indefinite time integrals in the definitions of $\zeta_{I2}^{(v)}$ and $\delta P_{\text{nad}\, 2}^{(v)}$. On the other hand, for the versions of $\zeta$ for which the threading was chosen with $\widetilde E=0$, or the original $\zeta_2$, the definitions include inverse Laplacians (see Eq. \eqref{z2}). In both cases, non-locality is present in some form, either in time or in space, and there is no version of the curvature perturbation which evades both of these issues. However, in both cases, the difficulties of the calculation are resolved by solving additional differential equations, both of which require boundary conditions. In the case of the inverse Laplacian, the equation to solve is a Poisson equation, which only depends on first order quantities at a single time, while for the case of the integrals in time, knowledge of the full time evolution of second order quantities is required ($\n^2J_2$ in Eq. \eqref{zI2v}, for example). This seems to render the quantities without integrals in time more amenable for situations that require the calculation of $\zeta$ from its definition, such as when its value is evaluated from the value of scalar field or density perturbations. In any case, all these issues disappear in the large scale approximation, for which the inverse Laplacian term in question has a well defined limit and the integrals vanish.

Moreover, we found that, on large scales, the
evolution of $\zeta$ is sourced by the transverse traceless part of the anisotropic stress tensor, as well as non-adiabatic pressure. Both quantities must therefore be negligible for any version of $\zeta$ to be
conserved. Furthermore, the version of the curvature perturbation which is exactly conserved is the one based on the determinant of $g^{ij}$ and comoving threading, $\zeta_{I}^{(v)}$, Eq. \eqref{zI2v}. Other definitions may evolve with the evolution of tensor and vector modes, should such an evolution be allowed by the theory of gravitation under study. For General Relativity, however, vector perturbations are usually very small and the evolution of tensor modes is negligible in the absence of anisotropic stress; therefore all versions of the curvature perturbation are approximately conserved on large scales.
%

The results presented here are valid as long as the energy and momentum conservation equations, Eqs. \eqref{evorho1}, \eqref{evov1} and \eqref{evorho2}, are satisfied. This will be true if the stress-energy tensor is covariantly conserved, i.e. $\n_\mu T^{\mu\nu}=0$, and the connection is the Levi-Civita connection (i.e. no torsion is present). This is the case in GR, but also in other theories, such as Massive Gravity and Bigravity \cite{D'Amico:2011jj,DeFelice:2014nja}. The latter theories are interesting in this context, as the tensor modes evolve differently due to the non-zero mass of the graviton \cite{Fasiello:2015csa} and therefore, $\zeta_D$ and $\zeta_I$ would be the only versions of the curvature perturbation that are conserved. 

Furthermore, the usefulness of these results may be extended to theories of gravity for which $\n_\mu T^{\mu\nu}\neq0$. This is possible if one can perform a conformal transformation to the Einstein frame and apply the same ideas to the effective stress-energy tensor that arises as the r.h.s. of the new field equations. The difference between our standard scenario and a modified one is that the effective matter quantities thus defined, would not have the same physical significance as the ones we use in this work. Therefore, in those modified situations it may be less trivial to clearly say when the curvature perturbation is conserved, as, e.g. the effective $\delta P_{\text{nad}}$ may not be negligible when the true matter perturbations are adiabatic. The same could apply to the anisotropic stress.

Previous results on the subject of conserved quantities have not included
anisotropic stress \cite{Enqvist:2006fs} and have either done the calculations fully in the
large scale approximation \cite{LMS} or used a different quantity \cite{LV,LV2,LV3}.

\section*{Acknowledgements}
PC is funded by a Queen Mary Principal's Research Studentship and by a Bolsa de Excel\^encia Acad\'{e}mica of the Funda\c{c}\~{a}o Eug\'{e}nio de Almeida, KAM is
supported, in part, by STFC grant ST/J001546/1.
The tensor algebra package xAct \cite{xAct}, as well as its
sub-package xPand \cite{xPert,xPand}, were used in the derivation of
many of the equations presented in this work.
The authors are grateful to David Mulryne, Adam Christopherson, Timothy Clifton, Julien Larena and Raquel Ribeiro for useful discussions.

\appendix
\section{Second order gauge transformations}\label{app1}

Here we present the second order gauge transformations for the quantities used in this paper, given by Eq. \eqref{gauge2} when applied to the metric and the stress energy tensor. We show only the transformations for the variables defined in the original convention of the metric, Eq. \eqref{gij}, as the others can be easily obtained by combining the rules given here with the first order ones in the main text. 

The transformations for the metric quantities are given by
\begin{align}
\label{psi2g}
&\wt\psi_2=\psi_2-\H\alpha_2-\tfrac14\X^i_{\ i}+\tfrac14\n^{-2}\X^{ij}_{\ \ ,ij}\,,\\
&\wt E_2=E_2+\beta_2+\tfrac34\n^{-2}\n^{-2}\X^{ij}_{\ \ ,ij}-\tfrac14\n^{-2}\X^{i}_{\ i}\,,
\end{align}
with $\X_{ij}$ given by (see Ref. \cite{MW2008} for more details)
\begin{align}
\label{Xijdef}
\X_{ij}\equiv &\ 
2\Big[\left(\H^2+\frac{a''}{a}\right)\alpha_1^2
+\H\left(\alpha_1\alpha_1'+\alpha_{1,k}\xi_{1}^{~k}
\right)\Big] \delta_{ij}\nonumber\\
&
+4\Big[\alpha_1\left(C_{1ij}'+2\H C_{1ij}\right)
+C_{1ij,k}\xi_{1}^{~k}+C_{1ik}\xi_{1~~,j}^{~k}
+C_{1kj}\xi_{1~~,i}^{~k}\Big]
+2\left(B_{1i}\alpha_{1,j}+B_{1j}\alpha_{1,i}\right)
\nonumber\\
&
+4\H\alpha_1\left( \xi_{1i,j}+\xi_{1j,i}\right)
-2\alpha_{1,i}\alpha_{1,j}+2\xi_{1k,i}\xi_{1~~,j}^{~k}
+\alpha_1\left( \xi_{1i,j}'+\xi_{1j,i}' \right)
+\left(\xi_{1i,jk}+\xi_{1j,ik}\right)\xi_{1}^{~k}
\nonumber\\
&+\xi_{1i,k}\xi_{1~~,j}^{~k}+\xi_{1j,k}\xi_{1~~,i}^{~k}
+\xi_{1i}'\alpha_{1,j}+\xi_{1j}'\alpha_{1,i}
\,.
\end{align}

The transformations for the fluid quantities are
\begin{align}
&\wt{\delta\rho_2}=\delta\rho_2+\alpha_2\rho'+\alpha_1\lb\rho''\alpha_1+\rho'\alpha_1'+2\delta \rho'\rb+\lb2\delta \rho+\rho'\alpha_1\rb_{,k}\lb\beta_1^{,k}+\gamma_1^k\rb\,,\\
&\wt{\delta P_2}=\delta P_2+\alpha_2P'+\alpha_1\lb P''\alpha_1+P'\alpha_1'+2\delta P'\rb+\lb2\delta P+P'\alpha_1\rb_{,k}\lb\beta_1^{,k}+\gamma_1^k\rb\,,\\
&\wt{v_2}=v_2-\beta_2'+\n^{-2}\Xv_{\ ,k}^{\,k}\,,
\end{align}
with 
\begin{align}
\label{defXvi}
\Xv_i
\equiv&\ 
\xi'_{1i}\left(2\phi_1+\alpha_1'+2\H\alpha_1\right)-\alpha_1\xi''_{1i}
\nonumber\\
&
-\xi_1^k\xi'_{1i,k}+\xi_1^{k\prime}\xi_{1i,k}
-2\alpha_1\left(v_{1i}'+\H v_{1i}\right)
+2v_{1i,k}\xi_1^k-2v_1^k\xi_{1i,k}\,.
\end{align}

\section{Large scale limit of the gauge invariant quantities}\label{app2}

In this appendix, we supply the expressions for the different versions of the curvature perturbation in the large scale approximation, i.e. when all spatial gradients are taken to be negligible.

The large scale limit of $\zeta_2$ and $\zeta_{T2}$ is given by
\begin{align}
\label{z2largescale}
\zeta_2=\zeta_{T2}=-\frac{\H}{\rho'}\dr_2+\frac{1}{\rho'^2}\lb 2\H^2+\H'-\H\frac{\rho''}{\rho'}\rb\dr_1^2+\frac{2\H}{\rho'^2}\dr_1\dr'_1\,,
\end{align}
while that for $\zeta_{D2}$ and $\zeta_{D2}^{(v)}$ is
\begin{align}
\label{z2Dlargescale}
\zeta_{D2}=\zeta_{D2}^{(v)}=-\frac{\H}{\rho'}\dr_2+\frac{1}{\rho'^2}\lb \H'-\H\frac{\rho''}{\rho'}\rb\dr_1^2+\frac{2\H}{\rho'^2}\dr_1\dr'_1-\frac23h_{1ij}h_1^{ij}\,,
\end{align}
and the limit of $\zeta_{I2}$ and $\zeta_{I2}^{(v)}$ is
\begin{align}
\label{z2Ilargescale}
&\zeta_{I2}=-\frac{\H}{\rho'}\dr_2+\frac{1}{\rho'^2}\lb \H'-\H\frac{\rho''}{\rho'}\rb\dr_1^2+\frac{2\H}{\rho'^2}\dr_1\dr'_1-\frac23h_{1ij}h_1^{ij}+\frac13 S_{1i} S_{1}^i\,,\\
&\zeta_{I2}^{(v)}=-\frac{\H}{\rho'}\dr_2+\frac{1}{\rho'^2}\lb \H'-\H\frac{\rho''}{\rho'}\rb\dr_1^2+\frac{2\H}{\rho'^2}\dr_1\dr'_1-\frac23h_{1ij}h_1^{ij}+\frac13 \W_{1i} \W_{1}^i\,.
\end{align}

These expressions agree with the $\delta N$ formalism, where comparison is possible (see Ref.~\cite{LMS}).

\section{On intrinsic and extrinsic curvature}\label{app3}

In this appendix, we aim to clarify the relation between the different definitions of $\psi$ and the perturbation to both the intrinsic and extrinsic curvature of hypersurfaces of constant time. 

We begin by looking at the intrinsic curvature scalar. It is given by
\be
^{(3)}R=R+R_{\mu\nu}n^\mu n^\nu-K^2+K^{\mu\nu}K_{\mu\nu}\,,
\ee
in which $R_{\mu\nu}$ and $R$ are the 4D Ricci tensor and scalar, respectively, $n^\mu$ is the unit normal to the hypersurface, $K_{\mu\nu}$ is the extrinsic curvature and $K$ is its trace. The latter are given by
\begin{align}
K_{\mu\nu}=-\frac12[\L_n\gamma]_{\mu\nu}\,,\ \ \ \ K=-\n_\mu n^\mu\,,
\end{align}
with $\gamma$ the induced metric, given by $\gamma_{\mu\nu}=g_{\mu\nu}+n_\mu n_\nu$. The normal, $n^\mu$, is perpendicular to all vectors in the tangent space of the hypersurface. It is therefore often convenient to choose coordinates such that $n_i=0$. However, this specific coordinate choice means that the usual gauge transformation rules are not obeyed, and for this reason, we shall also compute these curvature scalars using the 4-velocity, $u^\mu$, to define the spatial hypersurface. In the latter situation, we will denote quantities with the superscript $(u)$.

We now present the calculations of these quantities up to second order in cosmological perturbation theory. The intrinsic curvature scalar is found to be
\begin{align}
\delta^{(3)}R_1&=\frac{4}{a^2}\n^2\psi_1\,,\\
\delta{}^{(3)}R_2&=\frac{1}{a^2}\Big[
4\nabla^2\psi_2-8C_{1km,}^{~~~~m}C_{1~~,n}^{kn}
+6C_{1mn,}^{~~~~k}C^{mn}_{1~~,k} -2C^k_{1~k,n}C^{m~~n}_{1~m,}
\nonumber\\
& +8C_1^{mn}\left(
C_{1mn,~k}^{~~~~~k}+C^k_{1~k,mn}-C_{1mk,n}^{~~~~~~k}-C_{1kn,m}^{~~~~~~k}
\right)\nonumber\\
&  +4\left(C^k_{1~k,j} C^{jn}_{1~,n}
+C_{1jk,}^{~~~j}C^{m~~~k}_{1~m,} -C^k_{1~n,m}C^{mn}_{1~~,k}
\right)\Big]\,,
\end{align}
where $C_{ij}$ is the perturbation to the spatial part of the metric. The relation between $\psi$ and the intrinsic curvature is clear at first order, as they are related linearly. This is the reason why the perturbation $\psi$ is called the curvature perturbation. However, this is only true for the original version of $\psi$, as given by the definition \eqref{gij}, since all other definitions include a contribution from the metric potential $E$. In any case, at second order, this simple connection between the intrinsic curvature and $\psi$ is lost, as there is no simple relation between any of our definitions of the curvature perturbation and $\delta{}^{(3)}R_2$.

Performing the same calculation using the 4-velocity to define the spatial hypersurface, one finds instead a connection to the curvature perturbation on comoving gauge, $\mathcal{R}$, since the first order result for $^{(3)}R^{(u)}$ is\footnote{Note that this quantity has the expected gauge transformation properties, since, being $0$ at the background level (because of the assumption of flatness), the Stewart-Walker lemma \cite{Stewart:1974uz,Stewart:1990fm} dictates it to be gauge invariant at first order. Notice that this does not happen in the calculation with $n$.}
\be
\delta^{(3)}R_1^{(u)}=\frac{4}{a^2}\n^2\left[\psi_1-\H(v_1+B_1)\right]=\frac{4}{a^2}\n^2\mathcal{R}_1\,.
\ee
At second order, however, the result is no longer related to the second order comoving curvature perturbation $\mathcal{R}_2$ in a simple way, i.e. $\delta^{(3)}R_2^{(u)}\neq\frac{4}{a^2}\n^2\mathcal{R}_2$. This can be seen by evaluating $\delta^{(3)}R_2^{(u)}$ in comoving gauge ($v=B=v_V^i=0$) and comparing it with $\frac{4}{a^2}\n^2\psi_2$. In this gauge, the intrinsic curvature is given by
\begin{align}
\delta^{(3)}R_2^{(u)}=\delta^{(3)}R_2+S_1^iw_i+S_1^{i,j}w_{ij}\,,
\end{align}
where $w_i$ and $w_{ij}$ are linear functions of the metric potentials. It is clear that this is not equal to $\frac{4}{a^2}\n^2\psi_2$, as there are no further cancellations that would recover that result. Therefore, one must conclude that none of our definitions of $\psi$ has a straightforward interpretation as the perturbation to the intrinsic curvature at an order higher than first.

Moving now to the scalar extrinsic curvature, we start by noting that it is proportional to the local expansion $\n_\mu n^\mu$ (or $\n_\mu u^\mu$, when choosing the velocity 4-vector to define the spatial hypersurface). It is well known that the integral of the expansion along world lines, with respect to proper time $s$, can be used to define a local scale factor \cite{LV,LMS}.
This integral is defined as
\be
\alpha=\frac13 \int \n_\mu n^\mu ds=-\frac13 \int K ds\,,
\ee
and the local scale factor is given by $e^\alpha$. This interpretation is further supported by the fact that, at the background level, one has $\alpha'=\H$. At first order, one finds
\be
\delta\alpha_1'=-\psi_1'-\frac13\n^2(B_1-E_1')\,.
\ee
This variable has some similarity with our definition of $\psi_T$, but still has a contribution from $B$, which is not present in any of our versions of the curvature perturbation at first order. Turning now to the situation with $u$ as the normal vector, the first order result is
\be
\delta\alpha_1^{(u)\prime}=-\psi_1'+\frac13\n^2(v_1+E_1')\,.
\ee
While this is still not equal to any version of $\psi$ directly, $\delta\alpha_1^{(u)}$ is, in fact, equal to $\zeta_{1T}$, when the latter is evaluated using a uniform density slicing. Going to second order, we find
\begin{align}
\delta\alpha_2^{(u)\prime}=&-\psi_2'+\frac13\n^2(v_2+E_2')+\frac13\left(-4C_{1ij} C_1^{ij\prime}+2\phi_1\n^2v_1+2 (v_{V1}^i+v_{1}^{,i})\left(\phi+C^j_{1j}\right)_{,i}\right.\nonumber\\
&\left. +\left[(v_{V1}^i+v_1^{,i}+B_1^{,i}-S_1^{i})(v_{V1i}+v_{1,i}+B_{1,i}-S_{1i})\right]'\right)\\
& - 2 (v_{V1}^i+v_{1}^{,i})\left(-\psi_1+\frac13\n^2\int(E_1'+v_1)d\tau\right)_{,i}\nonumber\,.
\end{align}
Again, this variable is not equal to any version of $\psi$, but it becomes exactly $\psi_I$, when evaluated using a comoving threading ($v=v_V^i=0$). This is equivalent to saying that, by applying the same procedure to this quantity, one would obtain a gauge invariant quantity that is equal to $\zeta_{I2}^{(v)}$. This is not surprising, given the results of Refs. \cite{LV,LV2,LV3,Enqvist:2006fs}, which found similar evolution equations for gauge invariants defined from the expansion scalar, $\Theta=\n_\mu u^\mu$.

We conclude our exposition of this appendix by noting that, even though the connection between the intrinsic curvature and $\psi$ is lost at second order, it is still possible to find a definition of $\psi$ which closely matches the extrinsic curvature scalar for comoving hypersurfaces (i.e. space-like hypersurfaces which are normal to the velocity 4-vector), at least when evaluated in a particular gauge. The reason why the version of $\psi$ that resembles $K$ is the one arising from the determinant of $g^{ij}$ can be explained by a relation between the determinant of the metric and the covariant divergence of a 4-vector. This is given by
\be
\n_\mu u^\mu=\partial_\mu u^\mu+\Gamma^\mu_{\nu\mu}u^\nu=\partial_\mu u^\mu+u^\nu\partial_\nu \log\left(\sqrt{-g}\right)\,,
\ee
in which $g=\det[g_{\mu\nu}]$. Furthermore, it can be shown that $\det[g^{ij}]$ is related to $g$ by
\be
g=g_{00}\left(\det[g^{ij}]\right)^{-1}\,,
\ee
and thus the previous relation becomes
\be
\label{nablaU}
\n_\mu u^\mu=\partial_\mu u^\mu+u^\nu\partial_\nu \log\left(\sqrt{-g_{00}}\right)-u^\nu\partial_\nu \log\left(\sqrt{\det[g^{ij}]}\right)\,.
\ee
Choosing a comoving threading is equivalent to setting $u^i=0$ and in that case it is straightforward to show that $u^0=\sqrt{-g_{00}}$. This implies that the first two terms on the r.h.s. of Eq.~\eqref{nablaU} cancel, and one finds 
\be
\left(\n_\mu u^\mu\right)_{\text{com}}=-u^\nu\partial_\nu \log\left(\sqrt{\det[g^{ij}]}\right)=3\frac{d}{ds}\left(\log a+\psi_I\right)\,,
\ee
in which we substituted $\det[g^{ij}]$ by the definition of $\psi_I$. Equivalently, one has
\be
-\left(\frac13 \int K^{(u)} ds\right)_{\text{com}}=\log a+\psi_I\,.
\ee
This shows $\psi_I$ to be the perturbation to the integrated extrinsic curvature of comoving hypersurfaces when written using a comoving threading. This result is valid at all orders and provides a clear interpretation to this perturbation derived from the determinant of the spatial part of the inverse metric.

\bibliographystyle{iopart-num}
\bibliography{zeta2_more}

\end{document}